\documentclass[12pt]{iopart}

\usepackage{graphicx}
\usepackage{color}
\usepackage{mathrsfs}
\usepackage{bm}
\usepackage{bbm}
\usepackage{amsfonts}
\begin{document}

\title{The entanglement dynamics of bipartite quantum system: Towards  entanglement sudden death}
\author{Wei Cui$^{1,2}$, Zairong Xi$^{1 *}$ and Yu Pan$^{1,2}$}
\address{%
$^1$Key Laboratory of Systems and Control, Institute of Systems
Science, Academy of Mathematics and Systems Science, Chinese Academy
of Sciences, Beijing 100190, People's Republic of China

$^2$Graduate University of Chinese Academy of Sciences, Beijing
100039, People's Republic of China
}%
\ead{zrxi@iss.ac.cn}

\begin{abstract}
We investigate the entanglement dynamics of bipartite quantum system
between two qubits with the dissipative environment. We begin with
the standard Markovian master equation in the Lindblad form and the
initial state which is prepared in the extended Werner-like state:
$\rho^{\Phi}_{AB}(0)$. We examine the conditions for entanglement
sudden death (ESD) and calculate the corresponding ESD time by the
Wootters' concurrence. We observe that ESD is determined by the
parameters like the mean occupation number of the environment $N$,
amount of initial entanglement $\alpha$, and the purity $r$. For
 $N=0$, we get the
analytical expression of both ESD condition and ESD time. For $N>0$
we give a theoretical analysis that ESD always occurs, and simulate
the concurrence as a function of $\gamma_0t$ and one of the
parameters $N, \alpha$, and $r$.

\end{abstract}

\maketitle

\section{Introduction}
Entanglement is responsible for the most counterintuitive aspects of
quantum mechanics \cite{Bell}, and motivated many philosophical
discussions in the early days of quantum physics \cite{Einstein}.
Recently it has been regarded as a resource for quantum information
processing \cite{Nielsen}. In fact, entanglement is one of the key
ingredient for quantum teleportation
\cite{Bennett,Bennett2,Bouwmeester}, quantum cryptography
\cite{Ekert} and is believed to be the origin of the power of
quantum computers, etc. However, a quantum system used in quantum
information processing inevitably interacts with the surrounding
environments (or the thermal reservoirs), which takes the pure state
of the quantum system into a mixed state \cite{H.P.Breuer}. Thus, it
is an important subject analyzing the entanglement decay induced by
the unavoidable interaction of the interested systems with the
environment \cite{weiss}. In one-party quantum system, this process
is called decoherence, and various methods have been proposed to
reduce this unexpected effect \cite{Mintert,Brand,Cui,zhang}. In
multiparty systems with non-local quantum correlations much interest
has been arisen in the dynamics of entanglement. For example,
entanglement sudden death (ESD), which means that it  disappears at
finite time, was discovered by Yu and Eberly \cite{Yu1, Yu2}. It
differs remarkably from the single qubit coherent evolution. The
interesting phenomenon has been experimentally observed for
entanglement photon pairs \cite{Almeida} and atomic ensembles
\cite{Laurat}.

ESD puts a limitation on the time when entanglement must be
exploited. The evolution of the entanglement and ESD have been
analyzed and various interesting results obtained
\cite{wang,Dajka,Dajka2,Roos,Konrad,Ikram,Qasimi,Huang,Yu3,Huang2,Tahira}.
Ref. \cite{Ikram} investigated the time evolution of entanglement of
various entangled states of a two-qubit system. For four different
initial states they analyzed the entanglement sudden death
conditions and time. From the simulation they got the conclusion
that the ESD always exists except for the vacuum reservoir. However,
it is still a question when the ESD occurs and what ESD time is in
the general cases and universal initial states. The aim of this
paper is to discuss the above problems for the standard Markovian
master equation in the Lindblad form and the initial states, the
extended Werner-like states: $\rho^{\Phi}_{AB}(0)$.

The paper is organized as follows. We first introduce the Wootters'
concurrence \cite{Wootters} and the extended Werner-like states:
$\rho^{\Phi}_{AB}(0)$. In Section III we give a standard Markovian
  master equation \cite{Turchette, Maniscalco}. The master equation
  is equivalent to a first order  coupled differential equations when
   the initial state is  the extended Werner like state  $\rho^{\Phi}_{AB}(0)$.
    In sections IV, we analyze the ESD conditions and ESD
time for the initial state $\rho^{\Phi}_{AB}(0)$. Conclusions and
prospective views are given in Section V.

\section{Concurrence and initial states}
A useful measure of entanglement is the Wootters' concurrence
\cite{Wootters}. For a bipartite system described by the density
matrix $\rho$, the concurrence $\mathcal{C}(\rho)$ is
\begin{equation}
\mathcal{C}(\rho)=\max(0,\sqrt{\lambda_1}-\sqrt{\lambda_2}-\sqrt{\lambda_3}-\sqrt{\lambda_4}),
\end{equation}
 where $\lambda_1, \lambda_2, \lambda_3$, and $\lambda_4$ are the
 eigenvalues (with $\lambda_1$ the largest one) of the ``spin-flipped"
 density operator $\zeta$, and
\begin{equation}
\zeta=\rho(\sigma_y^A\otimes\sigma_y^B)\rho^{*}(\sigma_y^A\otimes\sigma_y^B),
\end{equation}
 where $\rho^{*}$ denotes the complex conjugate of $\rho$ and
 $\sigma_y$ is the usual Pauli matrix. $\mathcal{C}$ ranges in
 magnitude from 0 for a disentanglement state to 1 for a maximally entanglement state.

The general
 structure of an ``X" density matrix \cite{Yu1, Yu2} is as follows
\begin{equation}
\hat{\rho}=\left(\begin{array}{cccc}
x&0&0&v\\
0&y&u&0\\
0&u^*&z&0\\
v^*&0&0&w
\end{array}\right),
 \end{equation}
 with $x,y,z,w$ real positive and $u,v$ complex quantities. Such states are general enough to include states such as the Werner
 states, the Bell states, \emph{et al.}. A remarkable aspect of the ``X" states is that the initial ``X'' structure is
 maintained during the Lindblad master equation evolution. This particular form of the density matrix allows us to
 analytically express the concurrence at time $t$ as \cite{Yu1}
 \begin{equation}
 \mathcal{C}_{\rho}^{X}(t)=2\max\{0,|u|-\sqrt{xw},|v|-\sqrt{yz}\}.
 \end{equation}

In the present work, we will analyze in detail the two-qubit
entanglement dynamics in Markovian environment starting from the
initial ``X'' states defined in Eq. (3). We will examine the
exactly ESD time and entanglement evolution for two kind of
special states, the extended Werner-like states,

 \begin{equation}
\rho^{\Phi}_{AB}(0)=r|\Phi\rangle_{AB
AB}\langle\Phi|+\frac{1-r}{4}\emph{\textbf{I}}_{AB},
 \end{equation}
where $r$ the purity of the initial states,
$\emph{\textbf{I}}_{AB}$ the $4\times4$ identity matrix and
 \begin{equation}
\Phi_{AB}=(\cos(\alpha)|10\rangle+\sin(\alpha)|01\rangle)_{AB},
 \end{equation}
with $\alpha$ measuring the amount of initial entanglement. The
state $\rho^{\Phi}_{AB}(0)$ has the following form
 \begin{equation}
\rho_{AB}^{\Phi}(0) =\left(\begin{array}{cccc}
\frac{1-r}{4}&0&0&0\\
0&r\cos^2(\alpha)+\frac{1-r}{4}&r\sin(\alpha)\cos(\alpha)&0\\
0&r\sin(\alpha)\cos(\alpha)&r\sin^2(\alpha)+\frac{1-r}{4}&0\\
0&0&0&\frac{1-r}{4}
\end{array}\right).
 \end{equation}

\begin{figure*}
\centerline{\scalebox{1.2}[0.8]{\includegraphics{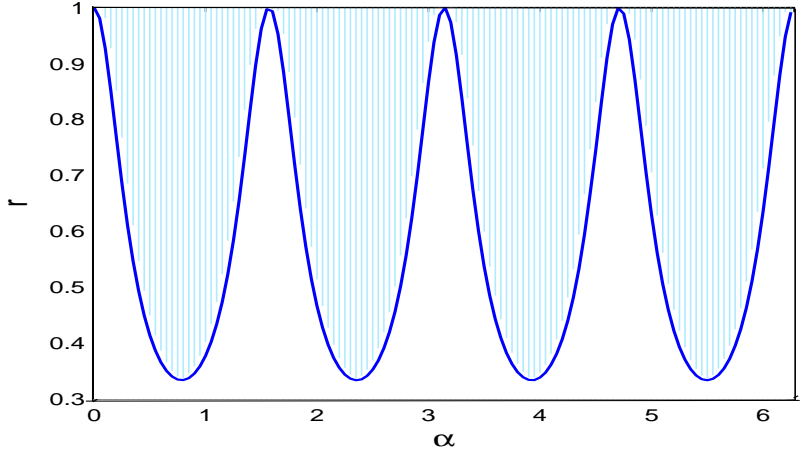}}}
\caption{(Color online)Plot the initial entangled area (colored) of
the extended Werner-like states for $\alpha\in[0, 2\pi]$. The blue
line is $r=\frac{1}{1+2|\sin(2\alpha)|}$.}
\end{figure*}

Obviously, the state in Eq. (5) is reduced to the standard Werner
state when $\alpha=\pi/4$, and  the Werner-like state become totally
mixed state for $r=0$, while  the well-known Bell state for $r=1$.
Note that, the extended Werner-like state $\rho^{\Phi}_{AB}(0)$
contain both separate state and entangled state. According to Peres'
criterion \cite{Peres}, when $\frac{1}{1+2|\sin2\alpha|}\leq r\leq1$
the extended Werner-like state would be entangled, otherwise it
would be separated. Fig.1 shows the entangled and separated areas.

\section{The master equation}

The standard two-qubit Markovian
 master equation is the following Lindblad form
 \cite{H.P.Breuer,Turchette,Maniscalco},
  \begin{equation}
  \fl
 \frac{d\rho}{dt}=\frac{\gamma_0(N+1)}{2}\sum_{j=1}^2\{2\sigma_j^{-}\rho\sigma_j^{+}-\sigma_j^{+}\sigma_j^{-}\rho-\rho\sigma_j^{+}\sigma_j^{-}\}\\
 +\frac{\gamma_0N}{2}\sum_{j=1}^2\{2\sigma_j^+\rho\sigma_j^--\sigma_j^-\sigma_j^+\rho-\rho\sigma_j^-\sigma_j^+\},
 \end{equation}
where $N=1/(e^{\frac{\hbar \omega_0}{K_BT}}-1)$, the mean occupation
number of the environment oscillators, $\gamma_0$ is the spontaneous
emission rate. As introduced before, we assume the initial state to
be a ``X'' state. When substituting (3) into the master equation (8)
we obtain the following first-order coupled differential
 equations,

 \begin{equation}
 \fl
\left(\begin{array}{ccc}
\dot{x}(t)\\
\dot{y}(t)\\
\dot{z}(t)\\
\dot{w}(t)
\end{array}\right)
=\left(\begin{array}{cccc}
-2\gamma_0(N+1)&\gamma_0N&\gamma_0N&0\\
\gamma_0(N+1)&-\gamma_0(2N+1)&0&\gamma_0N\\
\gamma_0(N+1)&0&-\gamma_0(2N+1)&\gamma_0N\\
0&\gamma_0(N+1)&\gamma_0(N+1)&-2\gamma_0N
\end{array}\right)
\left(\begin{array}{ccc}
x(t)\\
y(t)\\
z(t)\\
w(t)
\end{array}\right),
 \end{equation}
and
 \begin{equation}
  \begin{array}{rcl}
\dot{u}(t)&=&-(1+2N)\gamma_0u(t),\\
\dot{v}(t)&=&-(1+2N)\gamma_0v(t).
 \end{array}
 \end{equation}
The solution of the previous master equation can be found by
solving the system of differential equations. The reduced density
matrix elements $x(t), y(t), z(t), w(t), u(t)$ and $v(t)$ are
given

 (i) $N>0$
 \begin{equation}
  \begin{array}{rcl}
x(t)&=&c_1\frac{N}{N+1}-c_2\Gamma^2(t)-c_4\gamma_0N\Gamma(t),\\
y(t)&=&c_1+c_2\Gamma^2(t)+c_3\Gamma(t),\\
z(t)&=&c_1+c_2\Gamma^2(t)-c_3\Gamma(t)-c_4\gamma_0\Gamma(t),\\
w(t)&=&c_1\frac{N+1}{N}-c_2\Gamma^2(t)+c_4\gamma_0(N+1)\Gamma(t),\\
u(t)&=&c_5\Gamma(t),\\
v(t)&=&c_6\Gamma(t)

 \end{array}
 \end{equation}

(ii) $N=0$
\begin{equation}
  \begin{array}{rcl}
x(t)&=&d_4\Upsilon^2(t),\\
y(t)&=&d_2\Upsilon(t)+d_3\Upsilon(t)-d_4\Upsilon^2(t),\\
z(t)&=&-d_3\Upsilon(t)-d_4\Upsilon^2(t),\\
w(t)&=&d_1-d_2\Upsilon(t)+d_4\Upsilon^2(t),\\
u(t)&=&d_5\Upsilon(t),\\
v(t)&=&d_6\Upsilon(t)

 \end{array}
 \end{equation}
where $\Gamma(t)=e^{-(1+2N)\gamma_0t}$,
$\Upsilon(t)=e^{-\gamma_0t}$, and the coefficients $c_1, c_2, c_3,
c_4, c_5$,$c_6$ in Eq. (11), and $d_1, d_2, d_3, d_4, d_5$, $d_6$ in
Eq. (12) are determined by the corresponding initial conditions.

\section{Entanglement dynamics with the initial conditions}

\subsection{$N>0$}
 We now analyze  the entanglement dynamics. Starting from the initial states
$\rho_{AB}^{\Phi}(0)$ in Eq. (7),  the coefficients of Eq.(11) are
determined as

 \begin{equation}
  \begin{array}{rcl}
c_1&=&\frac{N(N+1)}{(2N+1)^2},\\
c_2&=&\frac{N(N+1)}{(2N+1)^2}-\frac{1-r}{4},\\
c_3&=&\frac{1}{2(2N+1)^2}+\frac{r}{2}(\cos^2(\alpha)-\sin^2(\alpha)),\\
c_4&=&-\frac{1}{\gamma_0(2N+1)^2},\\
c_5&=&r\sin(\alpha)\cos(\alpha),\\
c_6&=&0
 \end{array}
 \end{equation}
Then the concurrence of $\rho_{AB}^{\Phi}(t)$  is
\begin{equation}
C(\rho_{AB}^{\Phi}(t))=2\max\{0, |u(t)|-\sqrt{x(t)w(t)}\}.
\end{equation}
So the ESD appears when,
\begin{equation}
|u(t)|-\sqrt{x(t)w(t)}\leq0\Leftrightarrow u^2(t)-x(t)w(t)\leq0.
\end{equation}
Obviously,
$$t\rightarrow+\infty,~~~~u^2(t)-x(t)w(t)=-\frac{N^2(N+1)^2}{(2N+1)^4}<0,$$
and
\begin{equation}
t=0,~~~~~u^2(t)-x(t)w(t)=r^2\sin^2(\alpha)\cos^2(\alpha)-\frac{(1-r)^2}{16}.
\end{equation}
Thus, ESD always occurs if the initial state is entangled, which
means $\frac{1}{1+2|\sin2\alpha|}\leq r\leq1$, according to Peres'
criterion.

Note that  Ref. \cite{Ikram}  found that when the mean thermal
photon number is not zero, in the thermal reservoir the entanglement
sudden death always happen from the simulation. However here we get
the same result from the theoretic analysis. In Fig. 2, 3, and 4 we
simulate the concurrence as a function of $\gamma_0t$ and one of the
parameters $N, \alpha$, and $r$, respectively. Fig. 2 is the
concurrence $\mathcal{C}_{\rho}^{\Phi}(t)$ as function of
$\gamma_0t$ and $N$, fixing the purity $r=1$ and initial degree of
entanglement $\alpha=\frac{\pi}{4}$, the Bell-like states. It shows
that ESD time is affected by $N$. The smaller  $N$, the longer its
ESD time. In Fig. 3, we plot the concurrence
$\mathcal{C}_{\rho}^{\Phi}(t)$ as function of $\gamma_0t$ and
initial entanglement $\alpha$ when $r=1$. When
$\alpha=\frac{\pi}{4}$ or  $\alpha=\frac{3\pi}{4}$ the initial
states are reduced to Bell-like states. The ESD time is sensibly
affected by $\alpha$. Fig. 4 is the concurrence as function of
$\gamma_0t$ and purity $r$. As exemplified before, when $0\leq
r\leq\frac{1}{1+2|\sin(2\alpha)|}$ the initial state is separated,
so we choose $r$ from $1/3$ to $1$. It shows that  the larger the
purity $r$ the longer the ESD time.

Here we study the entanglement dynamics of bipartite quantum system
in the global environment effect. However, Ref. \cite{Yu3} had shown
that under the classical niose effect entanglement may experience a
sudden death process even if the local coherence of one
participating particle is well preserved and the other one decays to
zero asymptotically. How do the local thermal reservoirs influence
the entanglement dynamics? Up to my knowledge, the master equation
needs to be reconstructed and the local environment effect embodied
by the spectral density of the thermal reservoir $J(\omega,T)$. The
one-body decoherence dynamics was studied in \cite{Cui,Ferrer} under
the local environment effect. Whether does the multipartite
entanglement dynamics under the local thermal reservoirs hold under
the classical noise effect proved by Yu etc al \cite{Yu3}? We will
study it in our further work.

\begin{figure*}
\centerline{\scalebox{1}[1]{\includegraphics{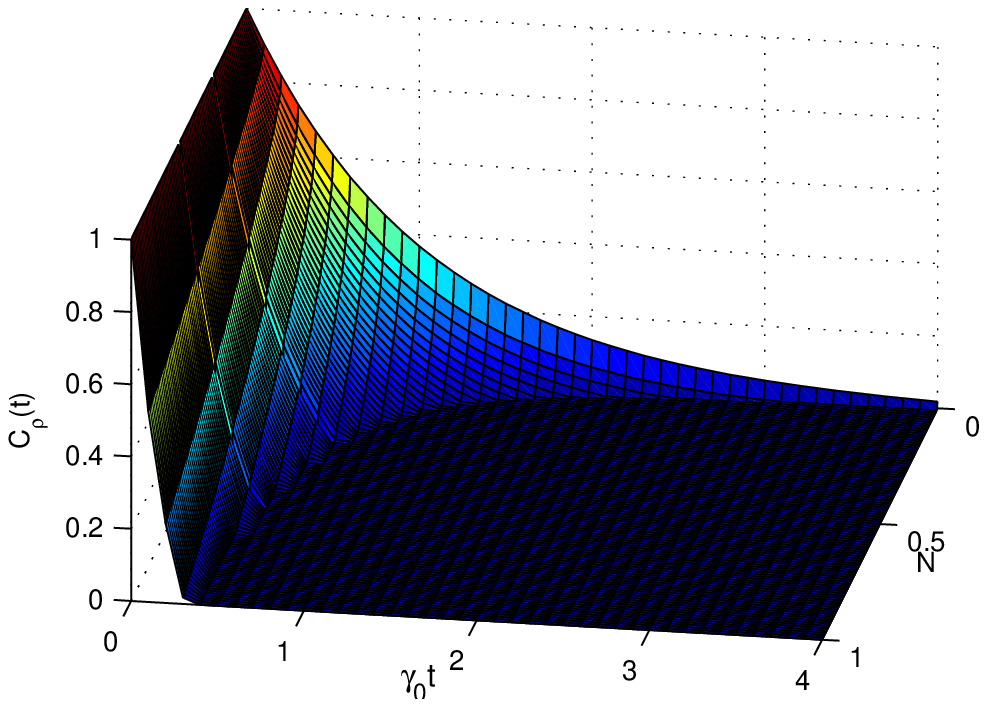}}}
\caption{(Color online)($r=1, \alpha=\pi/4$)Plot of concurrence of
$\mathcal{C}_\rho^{\Phi}(t)$ vs ``N" and $\gamma_0t$.}
\end{figure*}

\begin{figure*}
\centerline{\scalebox{1}[1]{\includegraphics{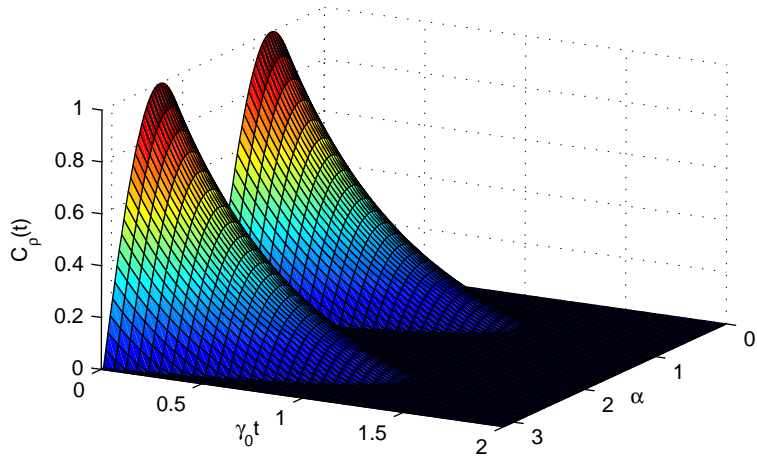}}}
\caption{(Color online)($r=1, N=0.1$)Plot of concurrence of
$\mathcal{C}_\rho^{\Phi}(t)$ vs $\alpha$ and $\gamma_0t$.}
\end{figure*}

\begin{figure*}
\centerline{\scalebox{1.2}[0.8]{\includegraphics{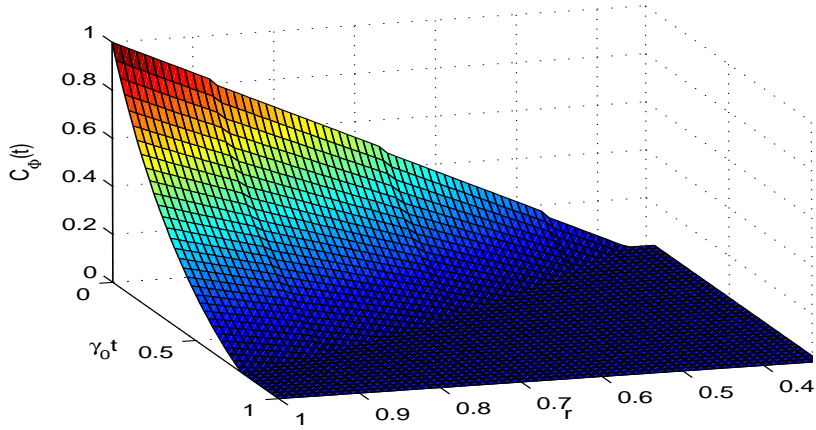}}}
\caption{(Color online)($\alpha=\pi/4, N=0.25$)Plot of concurrence
$\mathcal{C}_\rho^{\Phi}(t)$ vs ``r" and $\gamma_0t$.}
\end{figure*}

\subsection{$N=0$}
The coefficients in Eq. (12) have the form
\begin{equation}
  \begin{array}{rcl}
d_1&=&1,\\
d_2&=&1,\\
d_3&=&-\frac{1-r}{2}-r\sin^2(\alpha),\\
d_4&=&\frac{1-r}{4},\\
d_5&=&r\sin(\alpha)\cos(\alpha),\\
d_6&=&0.
 \end{array}
 \end{equation}
Thus
\begin{equation}
\fl
  \begin{array}{rcl}
&&u^2(t)-x(t)w(t)\leq 0\\
&\Leftrightarrow&\Upsilon^2(t)-\frac{4}{1-r}\Upsilon(t)+\frac{4}{1-r}-\frac{16r^2}{(1-r)^2}\sin^2(\alpha)\cos(\alpha)\geq0.
 \end{array}
 \end{equation}
So the  ESD occurs when,

\begin{equation}
\frac{1}{1+2|\sin(2\alpha)|}<r<\frac{-1+\sqrt{1+4\sin^2(2\alpha)}}{2\sin^2(2\alpha)},
\end{equation}
and the ESD time is
\begin{equation}
\gamma_0t^*=\ln(1-r)-\ln[2(1-\sqrt{r+r^2\sin^2(2\alpha)})].
\end{equation}
When $\gamma_0t\in[\gamma_0t^*,+\infty)$, the concurrence
$C^{\Phi}_{AB}=0$. In Fig. 5 we plot the ESD area. If  the initial
state is pure, i.e., $r=1$, we find that $x(t)\equiv0$. Then the
concurrence is
$C(\rho_{AB}^{\Phi}(t))=2|u(t)|=|\sin(2\alpha)|e^{-\gamma_0t}$,
which implies that the nonexistence of ESD. After we submitted our
paper, a closely related work appeared in \cite{Tahira}. Ikram et al
extended their former result \cite{Ikram} to more general state for
the zero temperature limit and provided some discussions on the
thermal case. Comparing with us, there are three main differences.
Firstly, although both papers studied the same system, a two 2-level
atom system, different initial states had been chosen. These two
initial states are independent, and the results are complementary.
Secondly, for pure 2-qubit entangled states in a thermal
environment, Ref. \cite{Tahira} says that ``We see in plots of
figure 3 that entanglement sudden death always happens for non-zero
average photon number in the two cavities and entanglement sudden
death time depends upon the initial preparation of the entangled
states." Here, a theoretical analysis is provided by Eqs. (14, 15,
16). Finally, in the vacuum reservoir we give the sufficient and
necessary condition Eq.(19) for the entanglement sudden death, and
answer the question when the ESD occurs and what ESD time is.
Furthermore, the ESD area relies on the initial purity $r$ and
initial entanglement $\alpha$, see Fig. 5. Here we give a more
detail examination and analysis.

The ESD condition and corresponding time for initial state
$\rho_{AB}^{\Phi}(0)$ at $N=0$ are summarized in the following
table.

\centerline{ \textbf{TABLE I. ESD condition and time for initial
state $\rho_{AB}^{\Phi}(0)$ at $N=0$}}

\begin{center}
\begin{tabular}{|c|c|}
\hline \multicolumn{1}{|c|}{$\alpha$}
 &\multicolumn{1}{c|}{(0,~$2\pi$)}\\
\hline

\textbf{Condition}&$\frac{1}{1+2|\sin(2\alpha)|}<r<\frac{-1+\sqrt{1+4\sin^2(2\alpha)}}{2\sin^2(2\alpha)}$\\
\hline

\textbf{Time}&$\gamma_0t^*=\ln(1-r)-\ln[2(1-\sqrt{r+r^2\sin^2(2\alpha)})]$  \\
\hline
\end{tabular}
\end{center}

\begin{figure*}
\centerline{\scalebox{1.2}[1]{\includegraphics{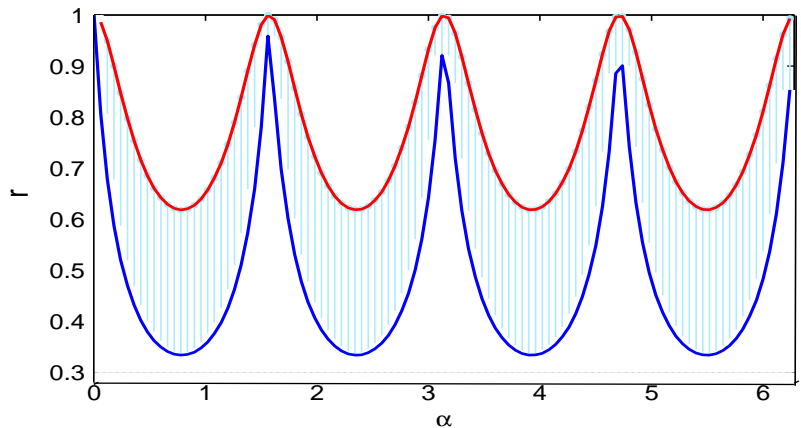}}}
\caption{(Color online)Plot the ESD area of  the extended
Werner-like states $\rho_{AB}^{\Phi}(t)$ with $N=0$ for
$\alpha\in[0, 2\pi]$. The red line is
$r=\frac{-1+\sqrt{1+4\sin^2(2\alpha)}}{2\sin^2(2\alpha)}$, and the
blue line is $r=\frac{1}{1+2|\sin(2\alpha)|}.$}
\end{figure*}

\section{Conclusions}
In summary, we have analyzed the interesting phenomenon of
entanglement sudden death determined by the dimensionless parameters
$\hbar\omega_0/k_BT, \alpha$, and $r$. Sufficient conditions for ESD
have been given for initial state $\rho^{\Phi}_{AB}(0)$. We examine
the conditions for ESD and calculate the corresponding ESD time by
the Wootters' concurrence. We observe that ESD is determined by the
parameters like $N$, amount of initial entanglement $\alpha$, and
the purity $r$. For
 $N=0$, we get the
analytical expressions of the ESD condition and ESD time. For $N>0$
we give a theoretical analysis that ESD always occurs, and simulates
the concurrence as a function of $\gamma_0t$ and one of the
parameters $N, \alpha$, and $r$. In fact, these above results can
also be obtain for other ``X" initial states, for example
$\rho^{\Psi}_{AB}(0)=r|\Psi\rangle_{AB
AB}\langle\Psi|+\frac{1-r}{4}\emph{\textbf{I}}_{AB},$ where $
\Psi_{AB}=(\cos(\alpha)|00\rangle+\sin(\alpha)|11\rangle)_{AB}.$

We analyze the dynamic behavior of entanglement in the open quantum
system But our ultimate aim is to control the entanglement such that
it be a resource for practical realization. Despite of the
noticeable progresses of entanglement, many fundamental difficulties
still remain. One of the problem is ESD due to the interactions
between system and environment; the other is that as the
$N$-particle increased, the entanglement becomes arbitrarily small,
and therefore useless as a resource \cite{Aolita}. Here we indicate
that preparing some initial states can help prolong the ESD time.
However, we think that it is a long way how to design some effective
control field to make the $N$-particle large enough and protect the
entanglement that make it practically useful.

\section*{Acknowledgments}
 We thank the referees for improving the manuscript. This research is supported by
the National Natural Science Foundation of China (No. 60774099, No.
60221301) and by the Chinese Academy of Sciences (KJCX3-SYW-S01).

\end{document}